\documentclass[usenatbib,twocolumn]{aastex631}

\usepackage{amsmath}
\usepackage{epsfig}
\usepackage{amssymb}
\usepackage{graphicx}
\usepackage{color}
\usepackage{natbib}
\begin{document}

% \received{July, 2024}
% \accepted{July, 2024}

\submitjournal{ApJ}

\shorttitle{Percolation in MTNG}
\shortauthors{Reg{\H o}s et al}
\title[Percolation in MTNG]{Percolation Statistics in the MillenniumTNG Simulations}

% These dates will be filled out by the publisher
\date{Accepted XXX. Received YYY; in original form ZZZ}

% Enter the current year, for the copyright statements etc.
% \pubyear{2024}

\author[0000-0002-9498-4957]{Enik\H o Reg\H os}
\affil{Konkoly Observatory, HUN-REN CSFK, Konkoly-Thege M. \'ut 15-17, Budapest, 1121, Hungary}
\affil{Magdalen College Oxford, UK}

\author[0000-0001-5976-4599]{Volker Springel}
\affil{Max-Planck-Institut f\"ur Astrophysik, Karl-Schwarzschild-Str. 1, D-85748, Garching, Germany}

\author{Sownak Bose}
\affil{Institute for Computational Cosmology, Department of Physics, Durham University, South Road, Durham, DH1 3LE, UK}

\author{Boryana Hadzhiyska}
\affil{Physics Division, Lawrence Berkeley National Laboratory, Berkeley, CA 94720, USA}

\author{C\'esar Hern\'andez-Aguayo}
\affil{Max-Planck-Institut f\"ur Astrophysik, Karl-Schwarzschild-Str. 1, D-85748, Garching, Germany}
\affil{Excellence Cluster ORIGINS, Boltzmannstrasse 2, D-85748 Garching, Germany}

\begin{abstract}
  The statistical analysis of cosmic large-scale structure is most often based on simple two-point summary statistics, like the power spectrum or the two-point correlation function of a sample of  galaxies or other types of tracers. In contrast, topological measures of clustering are also sensitive to higher-order correlations, and thus offer the prospect to access additional  information that may harbor important constraining power. We here revisit one such geometric measure of the cosmic web in the form of the so-called percolation analysis, using the recent MillenniumTNG simulation suite of the $\Lambda$CDM paradigm. We analyze continuum percolation statistics both for high resolution dark matter particle distributions, as well as for galaxy mock catalogues from a semi-analytic galaxy formation model within a periodic simulation volume of 3000 Mpc on a side. For comparison, we also investigate the percolation statistics of random particle sets and neutrino distributions with two different summed particle masses. We find that the percolation statistics of the dark matter distribution evolves strongly with redshift and thus clustering strength, yielding progressively lower percolation threshold towards later times. However, there is a sizable residual dependence on numerical resolution which we interpret as a residual influence of different levels of shot noise. This is corroborated by our analysis of galaxy mock catalogues whose results depend on sampling density more strongly than on galaxy selection criteria. While this limits the discriminative power of percolation statistics, our results suggest that it still remains useful as a complementary cosmological test when controlled for sampling density.
\end{abstract}

%\begin{keywords}
%\end{keywords}

\section{Introduction}

It is well known that non-linear cosmic structure formation via gravitational instability produces highly non-Gaussian structures today \citep{Peebles1974}, even if these structures grew from purely Gaussian fluctuations in the high-redshift initial density field. Testing whether the large-scale linear density field is really Gaussian and characterizing the degree of non-Gaussianity in the late-time Universe are therefore important tasks in cosmology. Yet the most commonly employed two-point statistics are completely insensitive to non-Gaussian correlations in the density field. It is however not straightforward to overcome this widely known fact. For example, using the full hierarchy of correlation functions through the sequence of three-point, four-point, and higher order $n$-point correlation functions is hampered by practical measurement difficulties, and a similar caveat applies to the bi-spectrum and other higher order measures in Fourier
space.

An alternative is given by integral measures of cosmic structure that adopt a more geometrical approach. Examples for this include the void probability function \citep{White1979}, Minkowski functionals \citep{Mecke1994}, Betty functions \citep{Pranav2017}, $k$-th nearest neighbor statistics \citep{Banerjee2021} -- or percolation statistics. In fact, historically, percolation theory has been advanced early on in seminal works by \citet{Zeldovich1982} and \citet{Shandarin1983} as a probe for cosmic structure formation. Percolation methods (sometimes called cluster analysis at the time) define a procedure to connect nearby cells or particles into local clusters via some kind of linking parameter. As the linking parameter is varied, the clusters slowly grow until a quite sudden transition occurs, at which point a global, long-range connectivity is first established for the biggest structure which  reaches opposite sides of the region under study, and subsequently starts to completely dominate all other clusters.

This phenomenon of percolation at a critical threshold of the linking parameter is connected to the theory of critical exponents and phase transitions \citep{Stauffer1979, Shklovskii1984}. The interest in studying the transition stems from its sensitivity to the detailed geometry of the point or density distribution, although it has been pointed out that not all distributions that are different can actually be readily distinguished by it \citep{Dekel1985}.

Over the recent decades, percolation has been used in numerous works as a probe for the cosmic web, in particular with the goal to characterize the geometry of large-scale structure, and to decide whether the cosmic web is more adequately described as, e.g.~cellular or filamentary. Among others, \citet{Einasto1984}, \citet{Postman1989} and \citet{Boerner1989} applied it to study superclusters, \citet{Klypin1993} used it to study galaxy clustering, and \citet{Yess1996} related it to the effective spectral index at the smoothing scale used in percolation analysis.  Galaxy surveys have also been analysed with percolation statistics, for example the IRAS 1.2 Jy survey \citep{Yess1997, Sathyaprakash1998}, the Las Campanas Redshift Survey \citep{Bharadwaj2000}, and the Sloan Digitial Sky Survey \citep{Pandey2005, Berlind2006}.

Insightful progress has been made by these studies which applied  percolation to the topological study of cosmological structures.  In particular, percolation has been shown to be both a quantitative measure of filamentary or cellular structures, and it helped to firmly establish that the galaxy distribution exhibits both planar and filamentary properties at the same time. Percolation has also been advanced as a discriminative tool between different cosmological scenarios for the formation of large scale structure, although this has seen only limited success thus far, perhaps a reason why the technique has been used less frequently lately in the literature, aside from a few notable exceptions \citep[e.g.][]{Zhang2018, Einasto2018, Busch2020}.

In this work, we revisit the use of percolation statistics as a measure of clustering, motivated by the much larger simulation volumes and galaxy numbers available in today's surveys. This significantly improved data quality may well usher in important differences to the studies done in the past. Thus, we want to clarify in this work whether percolation statistics should still be considered as an important higher-order complement to other measures of clustering, or whether one should rather focus on searching for more sensitive alternatives.

This work is structured as follows. In Section~\ref{SecMethodology}, we define the percolation statistics as we use it in this work, and we describe our simulations and our methodology for constructing galaxy mock catalogues. We then start out in Section~\ref{SecPercolationMatter} with an analysis of percolation in the dark matter component of the MTNG simulations at different numerical resolutions, and at different cosmological epochs. In Section~\ref{SecPercolationGalaxies}, we move on to investigate the percolation statistics of galaxy samples selected in different ways. We discuss our results in Section~\ref{SecDiscussion}, and give a summary of our conclusions in Section~\ref{SecConclusions}.

\begin{figure}
\resizebox{8.5cm}{!}{\includegraphics{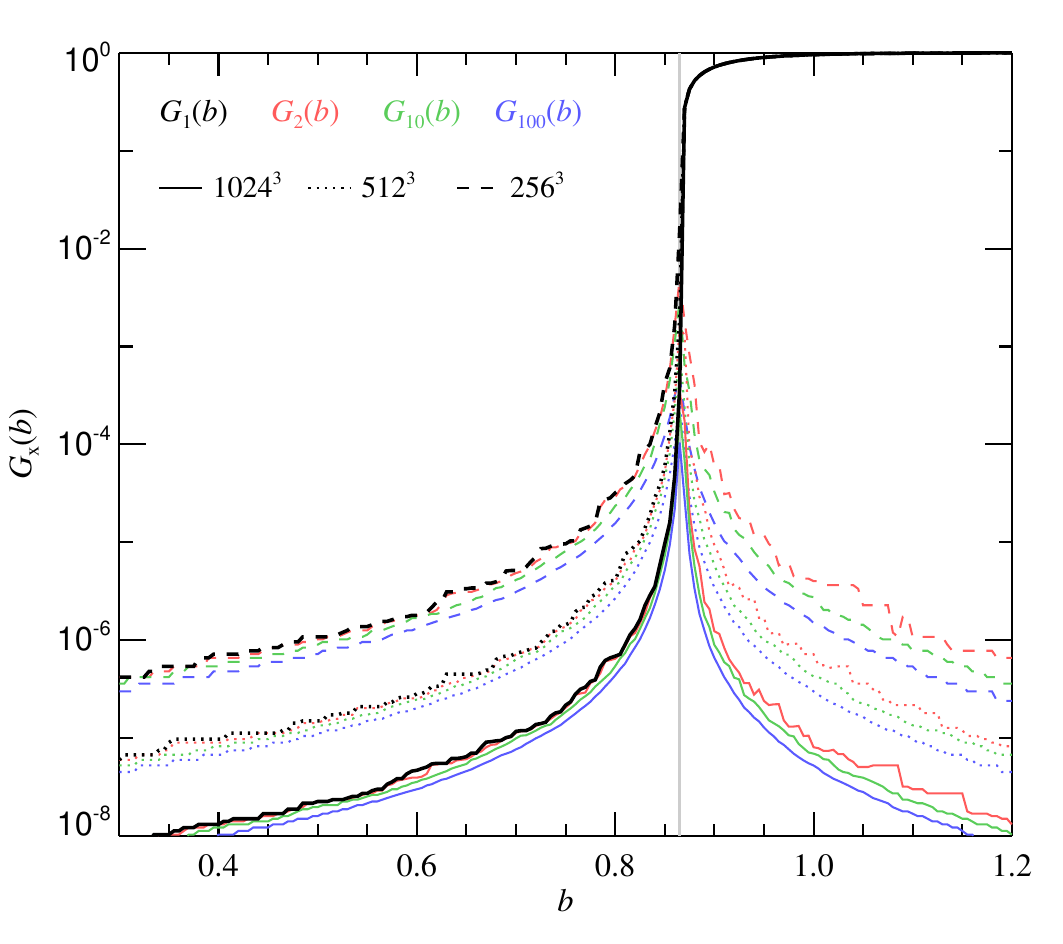}}
\caption{Percolation statistics for three different Poisson distributions of points, with $256^3$ (dashed), $512^3$ (dotted) and $1024^3$ (solid) particles. For all three cases, we show the mass of the biggest structure ($G_1$) in units of the total particle number, as well as the 2nd, 10th, and 100th largest structure as a function of the dimensionless linking length $b$. The percolation transition at $b\simeq 0.865$ is invariant with respect to the point density, as are the absolute particle numbers of the largest structures before the percolation transition occurs. However, the percolation curve depends on the particle number when the group sizes are expressed in terms of the fraction of mass they contain, as done here.
\label{FigRandomPointSet}}
\end{figure}

\section{Methodology}
\label{SecMethodology}

Percolation analysis is well known in other fields of physics and has
been introduced to the analysis of cosmic large-scale structure by \citet{Zeldovich1982}. Often it is carried out on a density field defined on a grid, in which case it is sometimes called site percolation as here the sites of a regular lattice are linked into the same cluster if they are adjacent and above (or below) a prescribed density threshold. In this variant of the percolation technique, cells of the same type are considered neighbors if they have a common face or edge. Neighbors of a neighbor are also considered neighbors, i.e.~the neighborhood criterion establishes equivalence classes, with sites in the same class forming a cluster. Sites in different clusters have no neighborhood relation with each other.

It is however also possible to work with discrete points instead, which leads to the concept of continuum percolation. We will here exclusively use the latter approach, which is more straightforwardly related to galaxy positions and/or dark matter particles. In particular, it sidesteps the introduction of technical nuisance parameters such as a finite grid resolution, which in turn necessitates the use of a (more or less arbitrary) smoothing scheme.

\subsection{Continuum percolation statistics}

For a set of $N_{\rm tot}$ points (e.g.~galaxy positions, halo centers, or dark matter particles) in a periodic box of volume $V= L^3$, we can define a mean particle spacing, 
\begin{equation}
d = (V/N_{\rm tot})^{1/3},
\end{equation}
and a linking length $l = b\, d$. This suggests the `Friends-of-Friends` provision that any two points with a spatial separation less than $l$ should be in the same group. The parameter $b$ thus defines a dimensionless linking length. In the analysis of cosmological N-body simulations, it is very common to use a dimensionless linking length of $b\simeq 0.2$ to identify groups of dark matter particles that constitute approximately virialized dark matter halos \citep{Davis1985}, which is known as the Friends-of-Friends (FOF) algorithm.

In contrast, in percolation analysis, ones asks the question how the size of the largest group varies when $b$ is systematically increased to even larger values \citep{Dekel1985}, up to a point where most of the points are contained in a single structure. This transition typically appears quite suddenly, i.e.~over a small variation of $b$ where the largest structure grows rapidly by a very large factor until it `percolates' and extends throughout the domain. For periodic boundary conditions, it then formally becomes infinitely big because the periodic images of groups of points are linked with each other. This qualitative change can be linked to the theory of phase transitions, and it is known to occur at different critical values of $b$ for point distributions that exhibit different morphological or topological distributions. Examination of the percolation transition  therefore offers the potential to discriminate between point distributions that are governed by different statistical processes.

As a quantitative measure of the percolation transition, we shall primarily employ the quantity
\begin{equation}
G_1(b) = N_1(b) / N_{\rm tot}, 
\end{equation}
which is the dimensionless size of the largest group. Here $N_1(b)$ is the number of points in the first ranked group for the dimensionless linking length $b$, and $N_{\rm tot}$ is the total particle number. It can also be of interest to look at the second most massive group,
\begin{equation}
G_2(b) = N_2(b) / N_{\rm tot},
\end{equation}
as well as at the $10^{\rm th}$ and $100^{\rm th}$ largest, $G_{10}(b)$ and $G_{100}(b)$, respectively, defined in an analogous way. We expect that $G_2(b)$ may decline for increasing $b$ after the percolation transition, because large groups should then be progressively linked to $G_1$, being replaced in the ranking by smaller groups. So the evolution of $G_2(b)$ is not necessarily monotonic when $b$ is increased, unlike that of $G_1(b)$. A similar consideration applies to $G_{10}(b)$ and $G_{100}(b)$, but we expect their evolution to be less noisy due to their reduced sensitivity to the size of just a single group. Note that the collection of $G_1(b)$, $G_2(b)$, $G_{10} (b)$ and $G_{100}(b)$ can also be viewed as probing 4 points on the cumulative halo mass function for a given linking length of $b$. The run of these four quantities with $b$ therefore probes how the massive end of the halo mass function varies with linking length.

In practice, we shall measure the percolation statistics with the FOF algorithm implemented in the {\small GADGET-4} \citep{Springel2021} cosmological code. This allows a fast and straightforward computation of $G_1(b)$ also for extremely large point sets. In practice, we vary $b$ between values of 0.3 and 1.2, in fine steps of $\Delta b = 0.005$.

\begin{figure}
\resizebox{8.5cm}{!}{\includegraphics{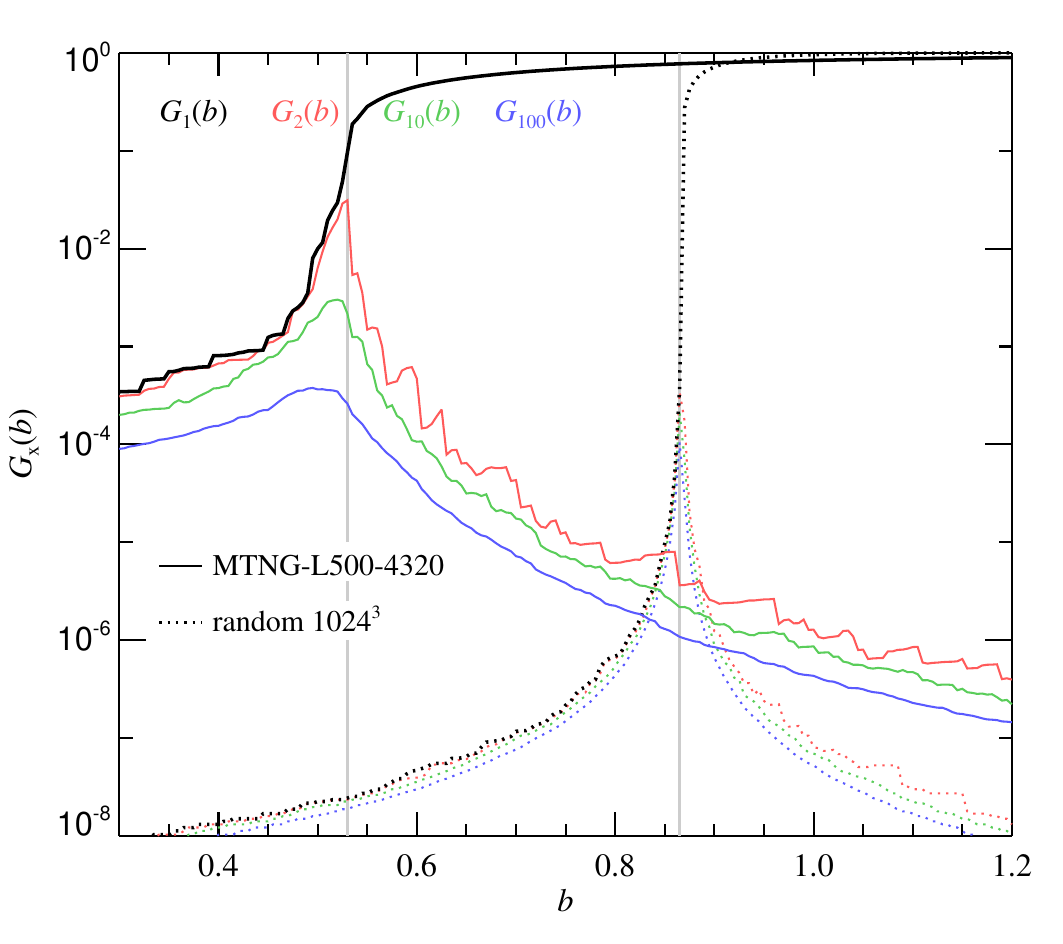}}
\caption{Dark matter percolation statistics for a high-resolution dark matter simulation at $z=0$ (solid lines), here the MTNG-DM-4320-A simulation with $4320^3$ particles in a $500\,h^{-1}{\rm Mpc}$ box size. For comparison, we also show the result for a Poisson distribution with $1024^3$ particles (dotted). Clearly, and not too surprisingly, the percolation statistics for the dark matter backbone of the cosmic web is dramatically different than for a random distribution. In particular, the percolation threshold is shifted to $b\simeq 0.53$, but also the mass fractions in the largest groups before and after percolation evolve in a fundamentally different fashion.
\label{FigHighResSim}}
\end{figure}

\begin{figure}
\resizebox{8.5cm}{!}{\includegraphics{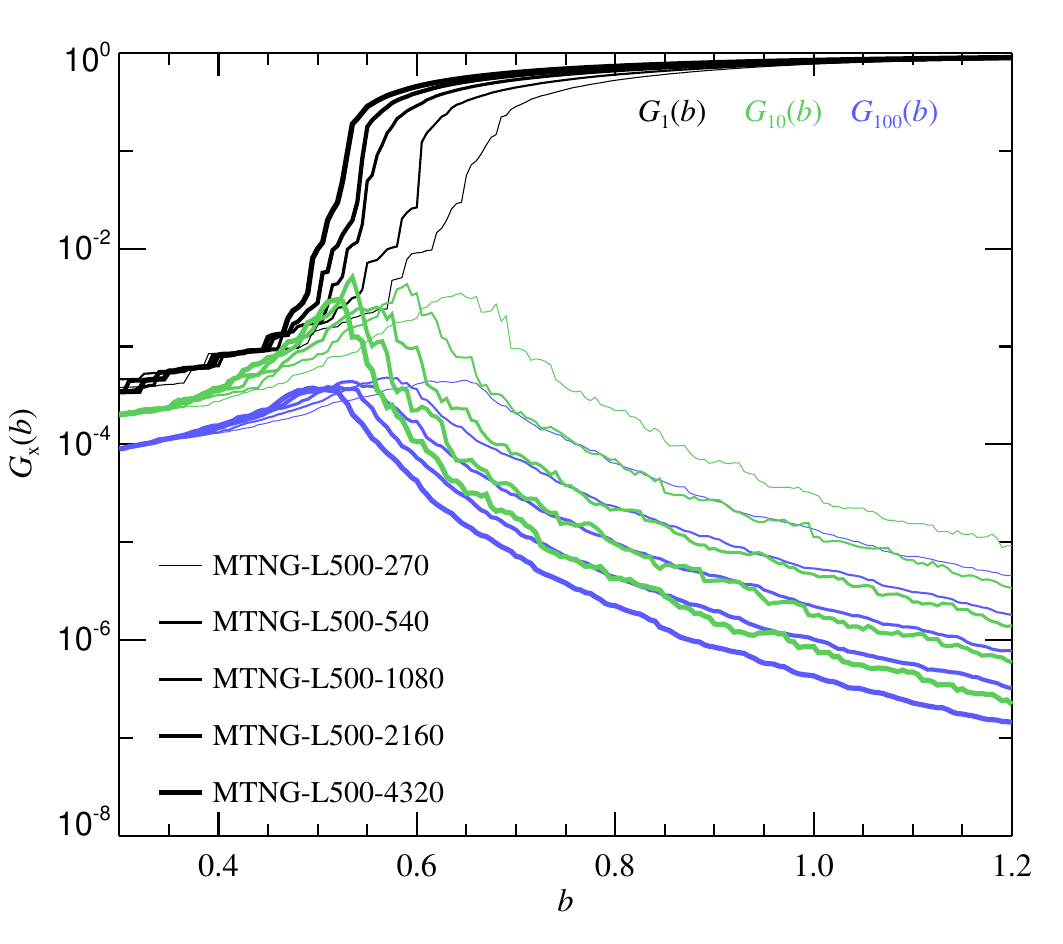}}
\caption{Convergence of the dark matter percolation statistics for N-body simulations at different numerical resolutions at $z=0$. We consider the `A'-series of dark matter only simulations in a $500\,h^{-1}{\rm Mpc}$ box carried out by the MTNG project. The line thickness systematically increases from $270^3$ to $4320^3$ particles, as labelled.
\label{FigDMDifferentResolutions}}
\end{figure}

\begin{figure*}
\resizebox{16.0cm}{!}{\includegraphics{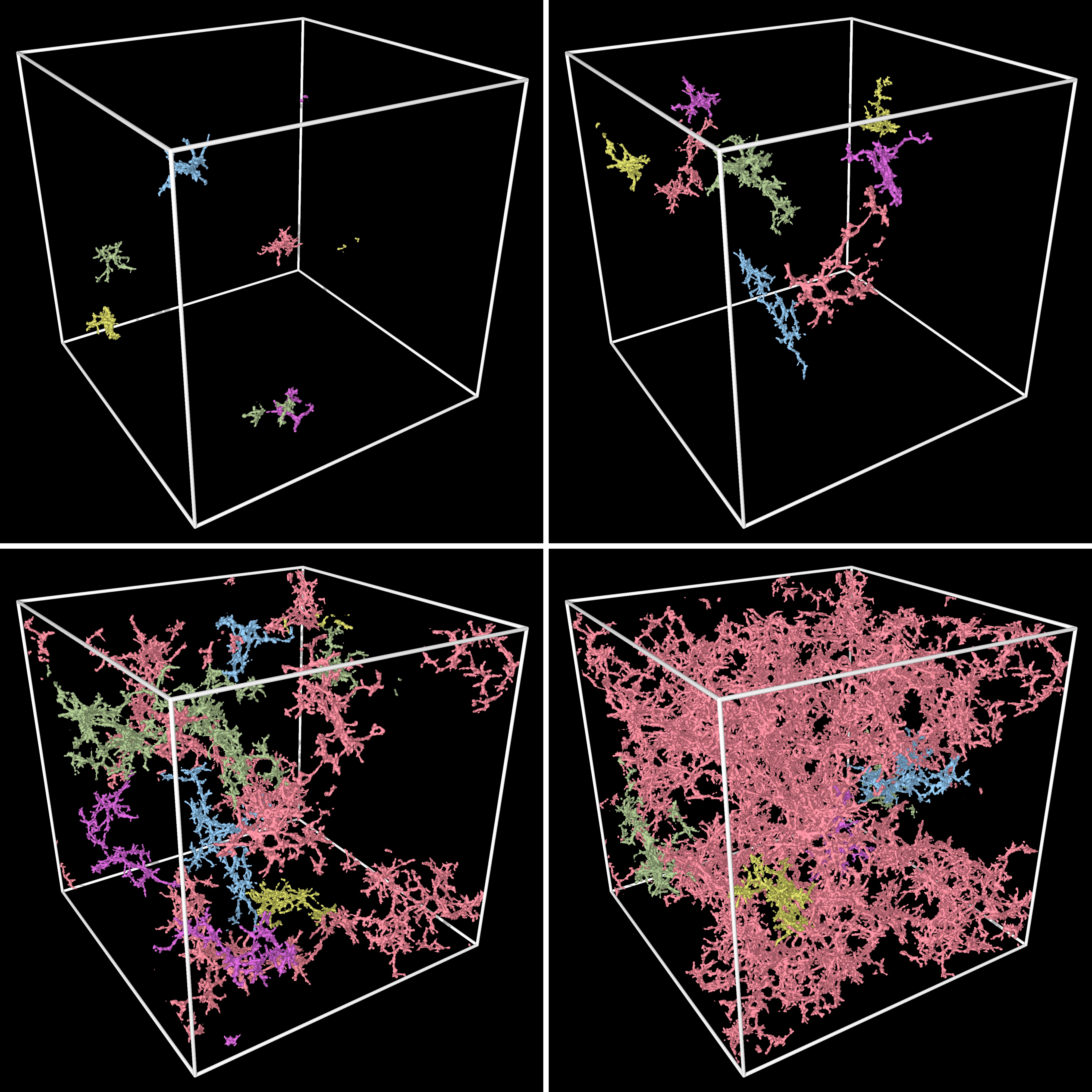}}
\caption{Visualization of the percolation statistics. From the top left to the bottom right panel, the linking length parameter increases from $b=0.50$ to $0.575$ in steps of $\Delta b = 0.025$. Each image visualizes the five largest groups identified in the MTNG-DM-1080-A simulation. For these linking lengths, the sizes of the largest group are 2.2, 6.5, 61.7 and finally 221.2 million particles, from top left to bottom right. These large groups resemble dendrites that connect clusters via their surrounding cosmic web filaments, until complete percolation has happened in the bottom right panel for the largest group. Further increases of $b$ in this case {\it reduce} the size of the groups, except for the largest one, because the latter tends to grow by merging with the second- or third ranked groups, which are then replaced in the ranking by smaller ones.
\label{FigVisualization}}
\end{figure*}

\subsection{The MillenniumTNG simulations and its mock catalogues}

The MillenniumTNG simulation project (or MTNG for short) makes theoretical predictions for the evolution of the Universe in sufficiently large volumes to link the development of large scale structure to non-linear galaxy formation. This is helpful to design precision tests of the $\Lambda$CDM cosmological paradigm using the upcoming new generation of extremely large galaxy surveys, such as Euclid or Rubin. The simulation suite of MTNG contains both fully hydrodynamical simulations in comparatively large volumes   as well as extremely large dark matter models in different box sizes, some of them also with massive neutrinos as hot dark matter admixture.  We refer to \citet{Aguayo2023} and \citet{Pakmor2023} for a comprehensive introduction of the simulations, and to the set of the other introductory papers on MTNG for further initial science results \citep{Bose2023, Kannan2023, Ferlito2023, Hadzhiyska2023a, Hadzhiyska2023b, Contreras2023, Delgado2023}.

In this work, we shall focus on the series of dark matter-only simulations within MTNG that use a $500\,h^{-1}{\rm Mpc} = 740\,{\rm Mpc}$ box size and feature particle numbers from $270^3$ to $4320^3$ particles, which is ideal for detailed convergence studies. We shall also use runs in a slightly smaller box of $430\,h^{-1}{\rm Mpc} = 632\,{\rm Mpc}$ that in addition to $2160^3$ dark matter particles include $540^3$ neutrino particles, either corresponding to a flavour-summed rest mass of $100\,{\rm meV}$ or  $300\,{\rm meV}$. Finally, we will consider very large semi-analytic galaxy catalogues at $z=0$ constructed with the techniques of \citet{Barrera2023} on top of the largest simulation in the MTNG suite, a model in a periodic box size of side length $2.04\,h^{-1}{\rm Gpc} = 3000\,{\rm Mpc}$ and dark mater particle number of  $N_{\rm cdm} = 10240^3$ as well as  $N_{\nu}= 2560^3$ additional neutrino particles. The galaxy properties are predicted with a new version of the {\small L-GALAXIES} code \citep{Springel2005} by following their evolution along the detailed halo and subhalo merger trees substructures constructed for the simulation. Note that these merger trees are quite large and information rich. They link together around $1.5\times 10^{11}$ gravitationally bound (sub)halos over 133 output times in more than 2 billion disjoint trees.

There are actually two realizations (A and B) for each of the dark matter simulations in MTNG, following the `fixed \& paired' technique of \citet{Angulo2016} for cosmic variance reduction. We will here restrict ourselves to using only the A-variant in the following\footnote{Actually B in the case of the mock galaxy catalogues, for a technical reason.}, as we find little if any difference between A and B for the percolation statistics.

\section{Percolation of the matter distribution}
\label{SecPercolationMatter}

\subsection{Random point sets}

It is interesting and instructive to first consider the percolation statistics of a random point set, i.e.~a Poisson distribution. In Figure~\ref{FigRandomPointSet}, we show results for three different point densities in a given box size, ranging from $256^3$ to $1024^3$. Clearly, the percolation transition is very sudden and rapid, and it occurs at a characteristic dimensionless linking length that is independent of the particle density. We measure $\overline b \simeq 0.865$ for this parameter. 

Note, however, that the full percolation curve is not invariant with respect to particle number \citep[unlike found by][]{Zhang2018}. This is because in the regime prior to percolation the largest group sizes tend to have similar absolute particle numbers for a given characteristic linking length. In the regime the groups do not yet know of the finite periodic domain they live in. However, after percolation, the evolution of the $G_1$ curve becomes universal, whereas the shrinking of the other group sizes stays non-universal due to a residual particle density dependence.

We shall consider the percolation threshold for a random distribution as an important reference point with which results for other particle distributions can be compared with. Note that it is not a priori clear how shot noise affects the percolation statistics, but it seems likely that any present amount of shot noise in the sampling of a field will tend to shift the percolation transition into the direction found here for random point distributions. Also note that shot noise cannot be easily subtracted or suppressed in the percolation statistics, unlike for the two-point correlation function or the power spectrum where its influence is analytically well understood.

\subsection{Cold dark matter}

We start our analysis of non-trivial point distributions with the dark matter of one of the high-resolution simulations of MTNG. We pick the $4320^3$ run in a $500\,h^{-1}{\rm Mpc}$ box, which has a particularly high mass resolution for a large volume cosmological simulation.

The result is shown in Figure~\ref{FigHighResSim}. Interestingly, this shows a dramatically different outcome than for a random point set. The percolation transition now occurs {\it much earlier}, already at $b\simeq 0.53$. In addition, the shapes of the percolation curves are dramatically different as well. At the lowest level, this of course establishes that the matter distribution in $\Lambda$CDM is extremely different from a Poisson distribution. This is expected of course, and thus it is arguably not a particularly high bar to detect this difference. But the percolation statistics does this in a very strong way. The compelling confirmation of the difference encourages therefore further analysis of this statistical measure.

An important question that arises right away, however, it to what extent this percolation transition is independent of numerical resolution (which it was, after all, for the Poisson distribution). To investigate this, we compare now in Figure~\ref{FigDMDifferentResolutions} the percolation statistics of a range of MTNG simulations, covering particle numbers from $270^3$ to $4320^3$, in steps of factors of eight. These simulations all have the same phases in the initial conditions, i.e.~they reproduce identical large-scale structure. This is also reflected in the good convergence of the percolation curves well before the percolation transition. This is because the largest structures have the same masses in the case when the linking length parameter has still moderate size. In this case, these structures correspond to halos, clusters, and superclusters.

However, at the percolation transition itself, this invariance with numerical resolution is lost, and higher resolution runs are found to systematically percolate earlier. With every step of a factor 8 better mass resolution, the residual shift in the percolation transition tends to become smaller, suggesting that the critical linking length is bounded by some lower limit which would be reached for infinite particle resolution.

At this point it is perhaps interesting to visually inspect the percolating structures, for the sake of developing a better intuitive understanding of their geometry. To this end we have extracted the 5 largest groups from our intermediate resolution run with $1080^3$ particles for four different $b$-values around the percolation transition, and show them in Figure~\ref{FigVisualization}. The structures are akin to the cosmic web backbone that is examined in various works that look at the topology of filaments and how they interconnect the massive halos at their intersections \citep[e.g.][]{Bond1996, Sousbie2011, Cautun2014}.

\begin{figure}
\resizebox{8.5cm}{!}{\includegraphics{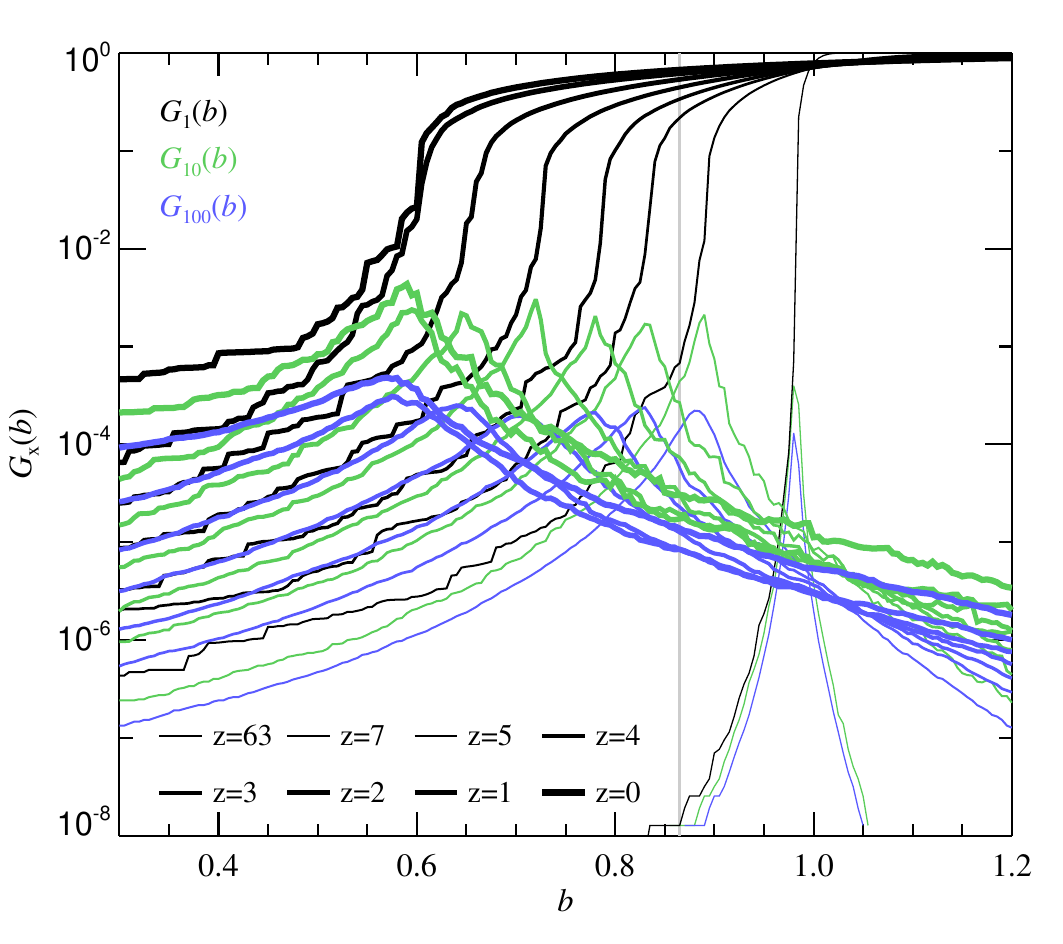}}
\caption{Redshift evolution of the dark matter percolation statistics for a typical N-body simulation, here illustrated for the MTNG-DM-540-A run of the MTNG project. For reference, the vertical grey line shows the location of the percolation transition for a random particle distribution.
\label{FigDMRedshiftEvolution}}
\end{figure}

Next we consider the time evolution of the percolation statistics of the cold dark matter distribution. As the amplitude of the clustering strongly evolves with time, this presumably is also reflected in the percolation transition, although how this manifests quantitatively is a piori not clear. In Figure~\ref{FigDMRedshiftEvolution} we therefore show measurements of the percolation statistics for a range of redshifts, in this case using the $540^3$ run, but the results for the other resolutions are qualitatively very similar.

Interestingly, at redshift $z=63$ (corresponding in fact to the initial conditions) we find an extremely high percolation threshold that lies only marginally below $b=1$, which is the mean particle separation itself. Clearly, this is because the particle distribution is here still close to the initial Cartesian grid used to represent the unperturbed Lagrangian particle distribution when constructing the initial conditions. In particular, this particle set is substantially sub-Poissonian, which is reflected in a percolation transition threshold that is actually larger than the one for a Poisson distribution.

With time, the percolation statistics is strongly evolving, reflecting the growth of cosmic structures. In particular, the percolation threshold evolves rapidly at high redshifts, and it crosses the one of the Poisson distribution at around redshift $z\simeq 6$. Interestingly, the evolution of the transition threshold slows down substantially at low redshift, potentially reflecting the beginning of structure formation freeze out in the $\Lambda$CDM cosmology. In any case, based on these results we clearly expect a strong $\sigma_8$-dependence of the percolation statistics when applied to the matter distribution of CDM cosmologies.

\begin{figure}
\resizebox{8.5cm}{!}{\includegraphics{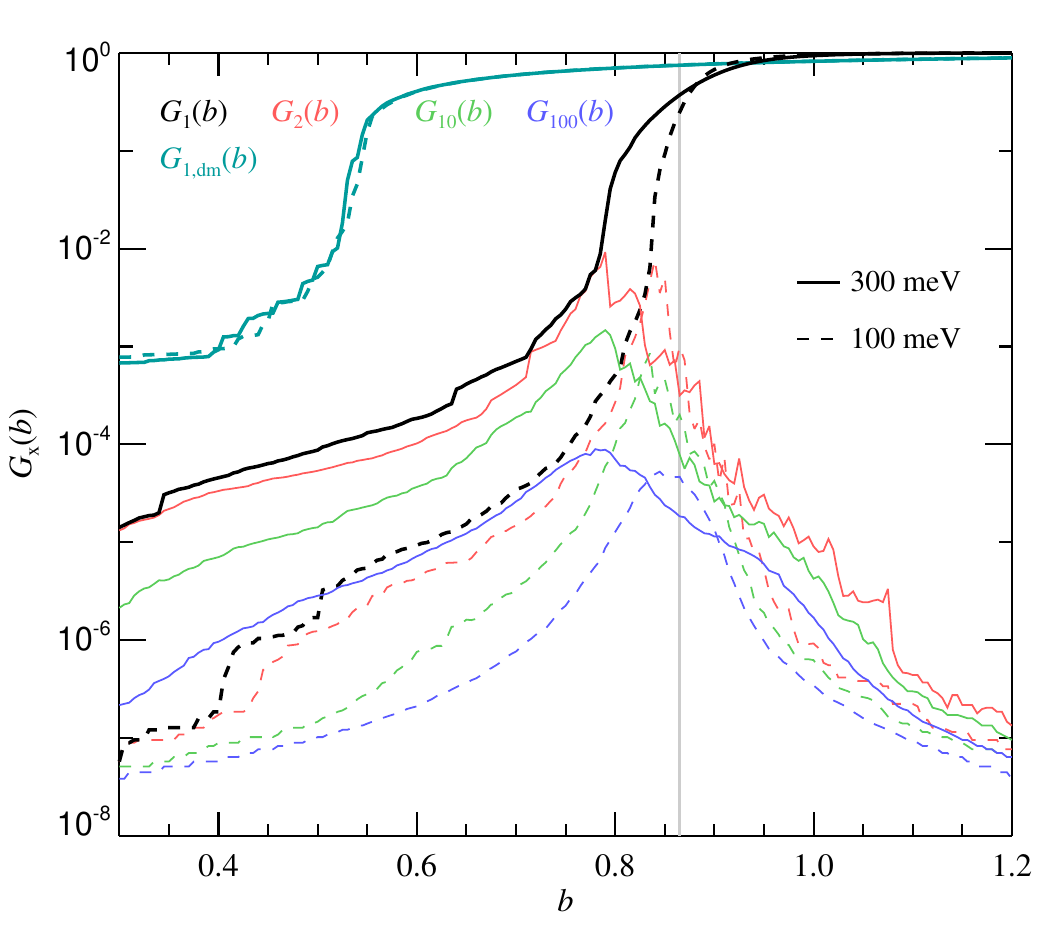}}
\caption{Percolation statistics for the neutrino particles in MTNG simulations with two different summed neutrino rest masses, as labelled. We also include measurements (upper left) for the corresponding dark matter particle distributions. For reference, the vertical grey line shows the location of the percolation transition for a random particle distribution.
\label{FigNeutrinos}}
\end{figure}

\subsection{Neutrinos as hot dark matter}

In this work we restrict ourselves to the cosmology of the MTNG simulations, i.e.~we do not investigate the cosmology dependence of the percolation statistics directly. However, since we also have simulations with massive neutrinos, we take a brief look at the neutrino distributions because this can be viewed as a hot dark matter admixture that can be expected to have a very different spatial distribution as the cold dark matter particles. This is because the neutrinos have very large, nearly relativistic velocities at high redshift. While they slow down with time due to their finite rest mass, they only respond late to the existing dark matter structures and start to cluster in and around the largest halos, once their velocities have become sufficiently small. Otherwise, their distribution will be nearly uniform. In our simulations, the neutrino particles faithfully sample this different velocity distribution  (albeit with a low sampling rate). In addition, we use the $\delta$f-technique of \citet{Elbers2021} to reduce shot noise from the neutrinos when their gravitational field is computed with the Poisson equation. 

We now test whether the percolation statistics reflects this expected difference in the spatial distribution of neutrinos relative to the dark matter. To this end we consider the neutrino particles in two of our smaller box simulations because here we have also runs available with different summed neutrino rest masses. 

In Figure~\ref{FigNeutrinos} we show the corresponding results, which interestingly show a substantial difference between the models with $300\,{\rm meV}$ and $100\,{\rm meV}$ neutrinos. The more massive neutrino model shows a substantially earlier percolation transition than the lighter model. The differences are consistent with the expectation that by $z=0$ the more massive neutrinos have begun to cluster more strongly than the lighter ones, which still move much faster. But even in the latter case, the neutrino distribution is clearly very different already from a Poisson distribution, demonstrating that even these light neutrinos have started to develop non-trivial clustering patterns at low redshift, and that this is not drowned in the substantial residual Poisson noise with which the neutrinos are represented in the MTNG N-body simulations. We also include in Fig.~\ref{FigNeutrinos} measurements of the percolation statistics for the corresponding dark matter particle distributions. We expect that the clustering of the dark matter in the simulation with the more massive neutrinos ends up being slightly stronger, consistent with the expectation that these neutrinos start to boost structure growth at late times compared with the lighter neutrinos that are still more smoothly distributed. Our percolation statistics results seem to indicate a very small systematic shift consistent with this expectation, but the effect is so small that it is not securely detected above the measurement noise.

\section{Percolation of the galaxy distribution}
\label{SecPercolationGalaxies}

We now turn to considering galaxy mock catalogues produced for our MTNG simulations \citep{Barrera2023}. This is in line with the originally intended application of percolation statistics as a means to quantify the three-dimensional geometry of the galaxy distribution, and to use this for a comparison between the observed cosmic large-scale structure and theoretical predictions. Ideally, such a comparison could be sensitive to features in the cosmic web morphology that are not differentiated by simple two-point clustering statistics. Note that unlike for the point sets analysed so far, the galaxy distribution can be affected by small-scale exclusion effects as galaxies cannot freely overlap due to their finite size. For the galaxy densities we will consider, their mean distance is far larger than their average size, so that the corresponding effects are weak.

For definiteness, we consider different number densities of galaxy samples, with values roughly in the ballpark of those of current surveys \citep[the same ones as][]{Angulo2014}. To implement these densities, we rank-order the original full galaxy catalogue for example by stellar mass, and then select the most massive galaxies down to a cut-off value that reproduces the desired space density. This procedure thus mimicks typical observational brightness or stellar mass cuts. Besides stellar mass, we also consider total halo mass of the parent halo the galaxy resides in, and the instantaneous star formation rate, as selection criteria. Especially for the SFR cut we expect that the resulting galaxy sample will differ substantially from the one defined through a stellar mass cut, as the latter will also contain many quenched massive galaxies that are disfavoured to be included in the SFR-selected catalogue. In the following, we are especially interested in the question how such expected differences show up in the percolation statistics.

\begin{figure}
\resizebox{8.5cm}{!}{\includegraphics{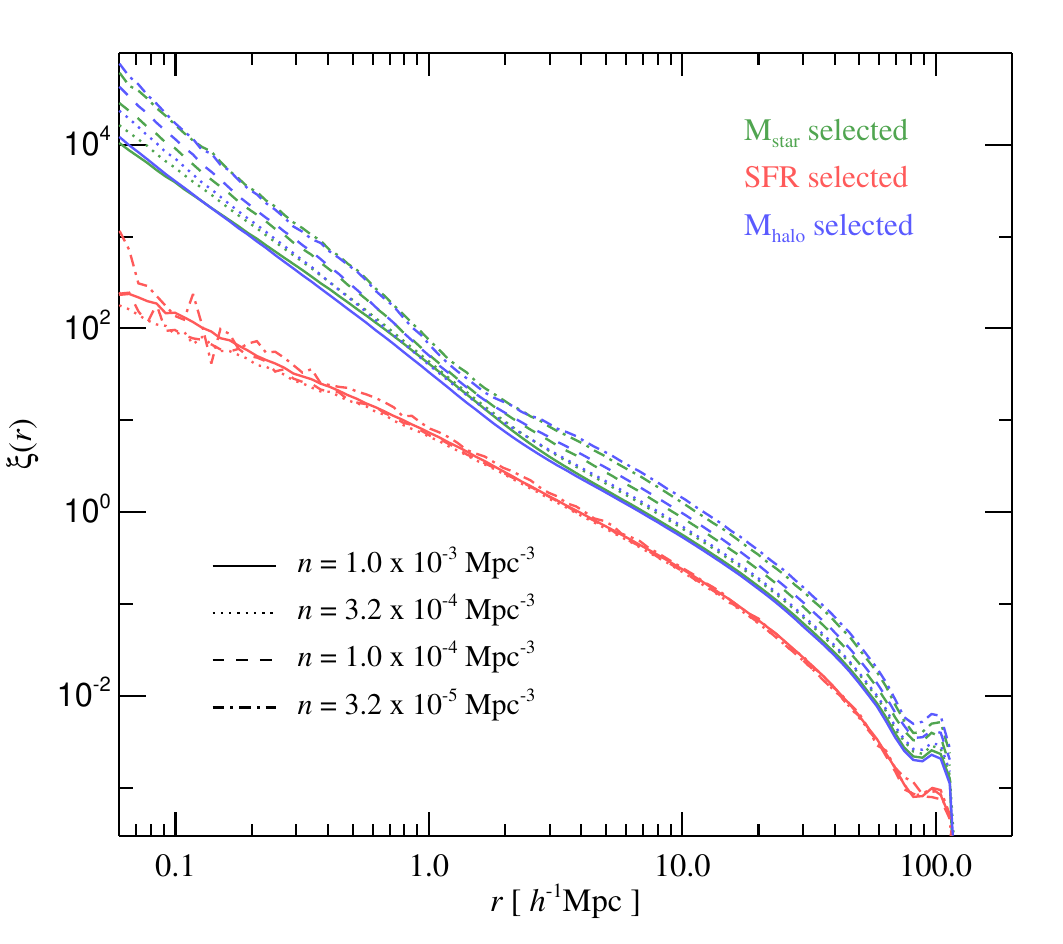}}
\caption{Two-point correlation function of the galaxy samples considered for our percolation analysis. We select either by stellar mass, halo mass, or current star formation rate, and consider four different survey densities, as labelled. These correlation functions are computed in real space. The bump at $\sim 120\,{\rm Mpc}$ is the baryonic acoustic oscillation feature, which is here well resolved thanks to the large 3000 Mpc box size of the underlying simulation model.
\label{FigTwoPoint}}
\end{figure}

Before we turn to this, it is instructive to first consider the ordinary two-point correlation functions of the different galaxy samples we have defined.  They are shown in Figure~\ref{FigTwoPoint} for the 4 number densities and the 3 selection criteria we consider. The  galaxy-two point correlation functions exhibit the familiar shape. Notably, we are able to detect the baryonic acoustic peak at $\sim 120\,{\rm Mpc}$ thanks to the large box size of our simulation. Importantly, the correlation function amplitude of the SFR-selected sample is considerably smaller than that of the stellar mass or halo mass selected samples. Furthermore, the clustering of the different SFR-selected galaxy samples is fairly similar, demonstrating in this case insensitivity to the number density of the considered galaxy sample. In contrast, the clustering amplitude of galaxies becomes systematically higher if rarer samples are selected by stellar or halo mass. This is due to the well-known effect that the galaxy bias increases for more massive halos \citep{Mo1996}. Note that for our mock catalogues, the two-point correlation functions are nearly identical for halo-mass selection or stellar-mass selection. For simplicity, we thus focus in the following on the comparison between halo mass selected and star formation rate selected percolation results.

\begin{figure}
\resizebox{8.5cm}{!}{\includegraphics{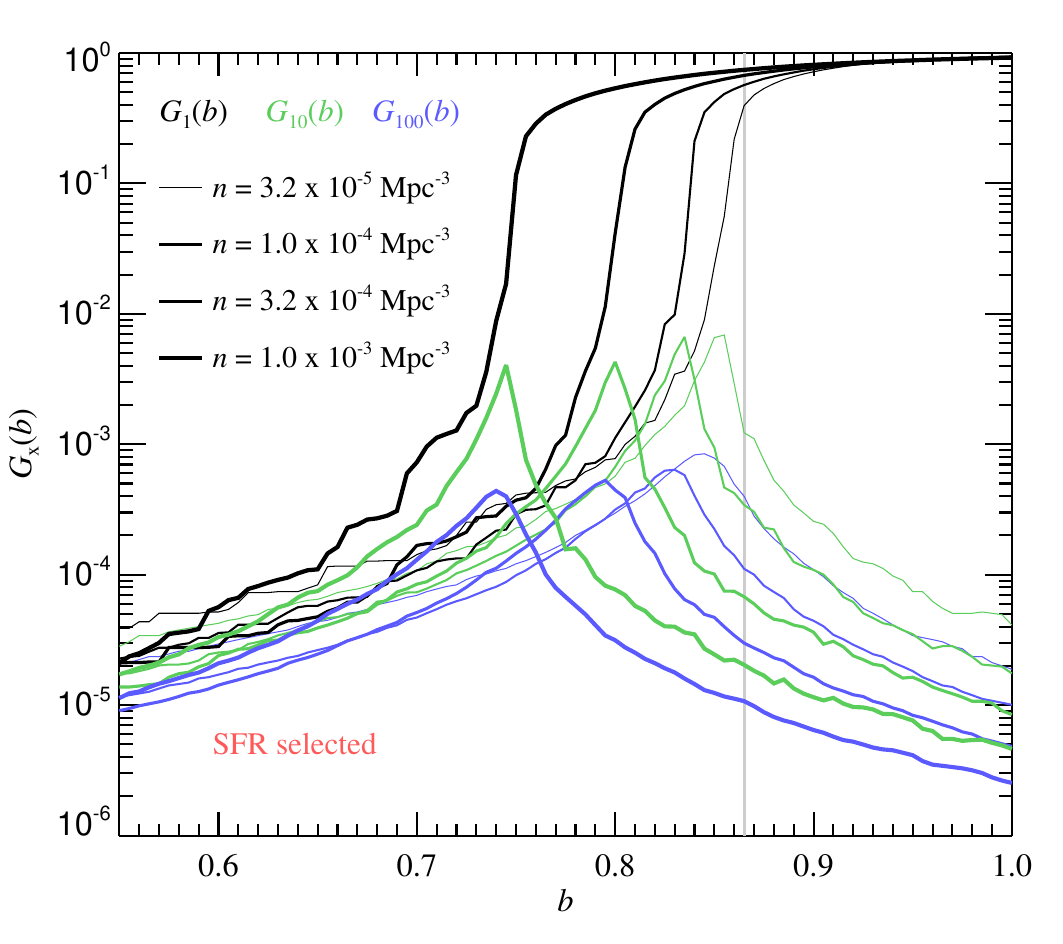}}
\caption{Percolation analysis for galaxy samples selected according to star formation rate down to different target space densities. Note that despite having nearly identical two-point correlation functions, the percolation threshold is markedly different for these four galaxy samples. For reference, the vertical grey line shows the location of the percolation transition for a random particle distribution.}
\label{FigGalSFRdifferentN}
\end{figure}

\begin{figure}
\resizebox{8.5cm}{!}{\includegraphics{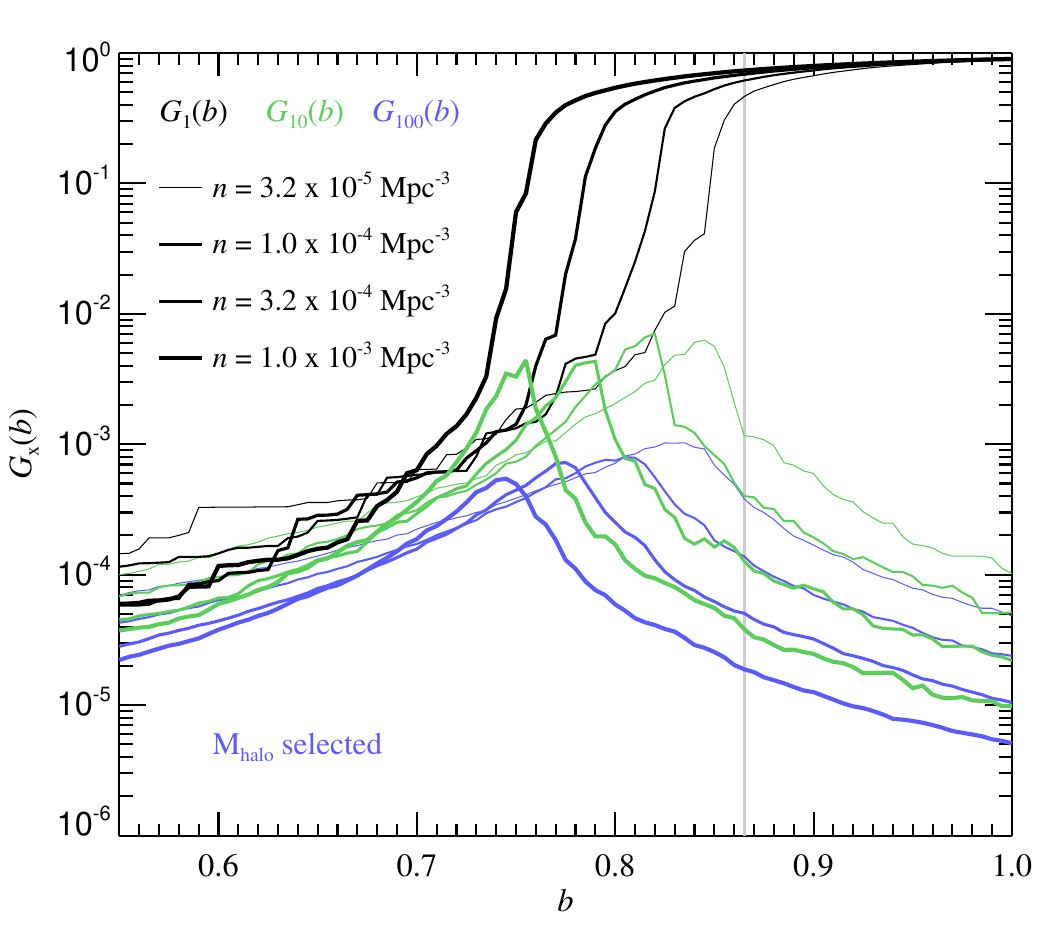}}
\caption{Percolation analysis for galaxy samples selected according to halo mass to different target space densities. We find significant drifts of the percolation threshold, in part driven by the bias evolution for these samples, in part driven by the sampling density. For reference, the vertical grey line shows the location of the percolation transition for a random particle distribution.
\label{FigGalMhalodifferentN}}
\end{figure}

We begin in Figure~\ref{FigGalSFRdifferentN} with the correlation statistics of galaxies selected by star formation rate, comparing the different sample number densities we have defined. Interestingly, the percolation threshold moves significantly with number density, from a value close to 0.75, to something around 0.85, quite close to the value one would get for a Poisson distribution. This is somewhat disconcerting as the two-point correlation functions of the SFR samples are nearly identical. Note, however, that the two-point correlation function is essentially free of shot noise (as this is compressed to a contribution at zero lag that is outside the radial range of interest), whereas the percolation statistics is affected by shot noise in a way that cannot be straightforwardly eliminated. The large differences in the results of Fig.~\ref{FigGalSFRdifferentN} seen for different galaxy number densities are unlikely caused by variations in the high-order correlations between the different samples, rather it seems more plausible that the shifts primarily reflect the growing influence of shot noise for the lower density galaxy samples.

\begin{figure*}
\resizebox{8.5cm}{!}{\includegraphics{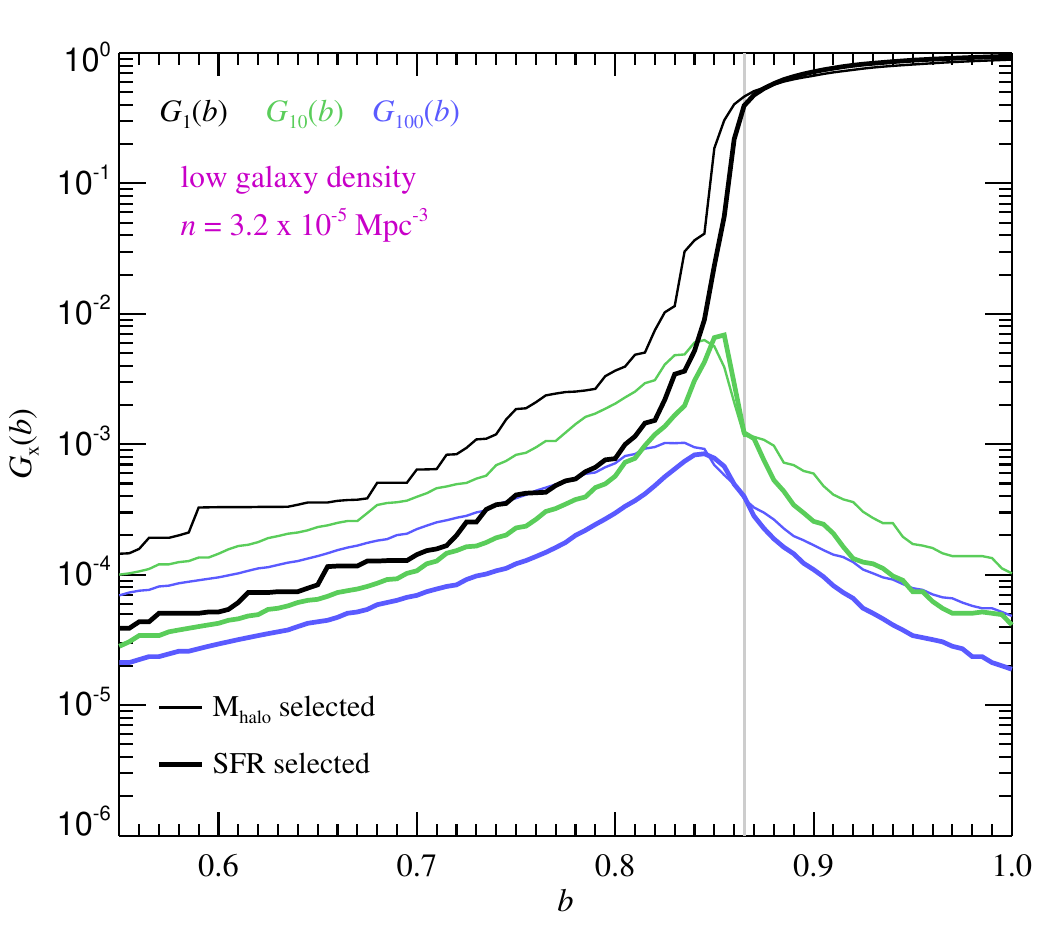}}%
\resizebox{8.5cm}{!}{\includegraphics{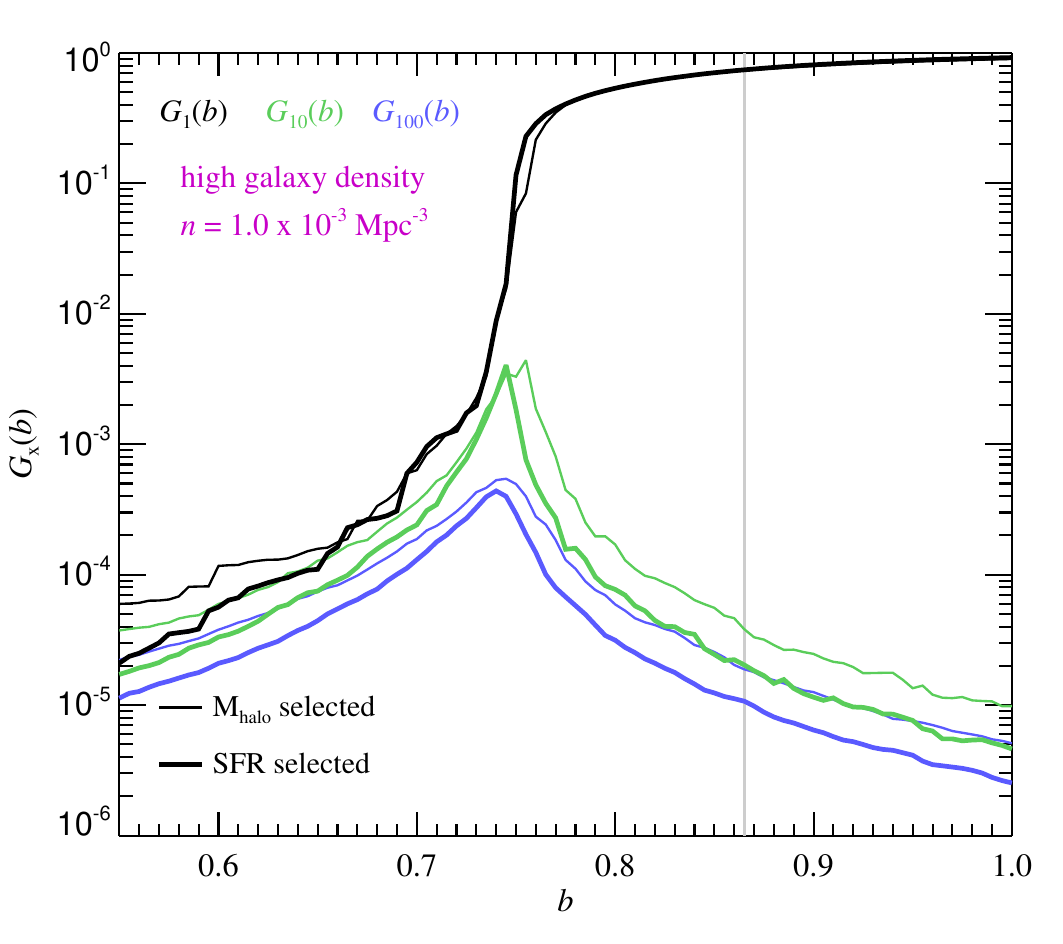}}%
\caption{Direct comparison of percolation statistics for different galaxy samples, selected at low (left panel) or high (right panel) galaxy space density, either through star formation or halo mass, as labelled. For reference, the vertical grey lines show the location of the percolation transition for a random particle distribution.
\label{FigCompMhalpSFR}}
\end{figure*}

This suspicion is corroborated by the corresponding results for galaxy samples selected by halo mass, which are shown in Figure~\ref{FigGalMhalodifferentN}. Here we again see a shift of the percolation threshold with galaxy number density. As the clustering amplitude increases towards the lower density samples, this should have a tendency to increase the two-point clustering signal, and thus reduce the percolation threshold. On the other hand, a lower number of galaxies increases the shot-noise, which we expect to have an opposite effect. The results of Fig.~\ref{FigGalMhalodifferentN} indicate that the shot-noise clearly wins and dominates over the differences from a modified galaxy bias.

This conclusion can be further strengthened by looking at a direct comparison of the percolation statistics of SFR- and halo-selected galaxy samples at the same tracer number density. We do this in Figure~\ref{FigCompMhalpSFR}, where we show in the left panel this comparison at our lowest number density and in the right panel at the highest number density we considered. The low density samples show a small but still clearly noticeable difference in their percolation statistics, reflecting also the quite substantial difference in their corresponding two-point clustering statistics. The percolation results for the high-density samples shown in the right panel are almost identical, however. This raises some serious questions about the constraining power of percolation statistics, because the corresponding galaxy samples can be easily told apart by their two-point clustering statistics. Note, however, that this does not necessarily mean that percolation statistics are not useful. If, for example, it was insensitive to the two-point correlation function but exclusively sensitive to higher-order correlation functions, the outcome we found might even be naturally expected.

Finally, we note that so far we have carried out our analysis in real space, for the sake of theoretical simplicity. In practice, of course, observed large-scale galaxy distributions will be in redshift space. We therefore briefly consider what changes are induced, if any, in the percolation statistics when one moves from real- to redshift space. To this end we shift our galaxy catalogues to redshift space by selecting one dimension as line-of-sight direction and shifting the galaxy coordinates to account for the peculiar velocity in that direction \citep[by adding it to that coordinate in real space to generate the corresponding redshift space coordinate, as in equation 8 of][]{Aguayo2023}. 

\begin{figure*}
\resizebox{8.5cm}{!}{\includegraphics{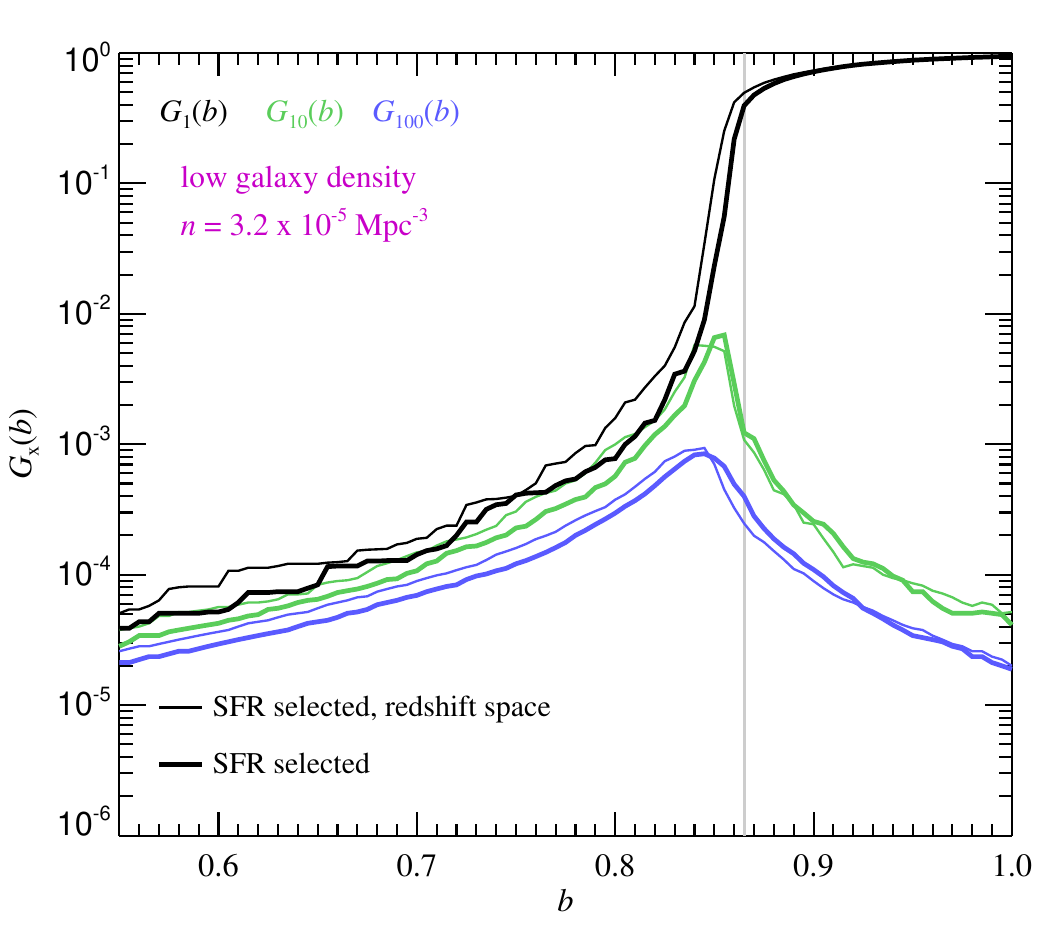}}%
\resizebox{8.5cm}{!}{\includegraphics{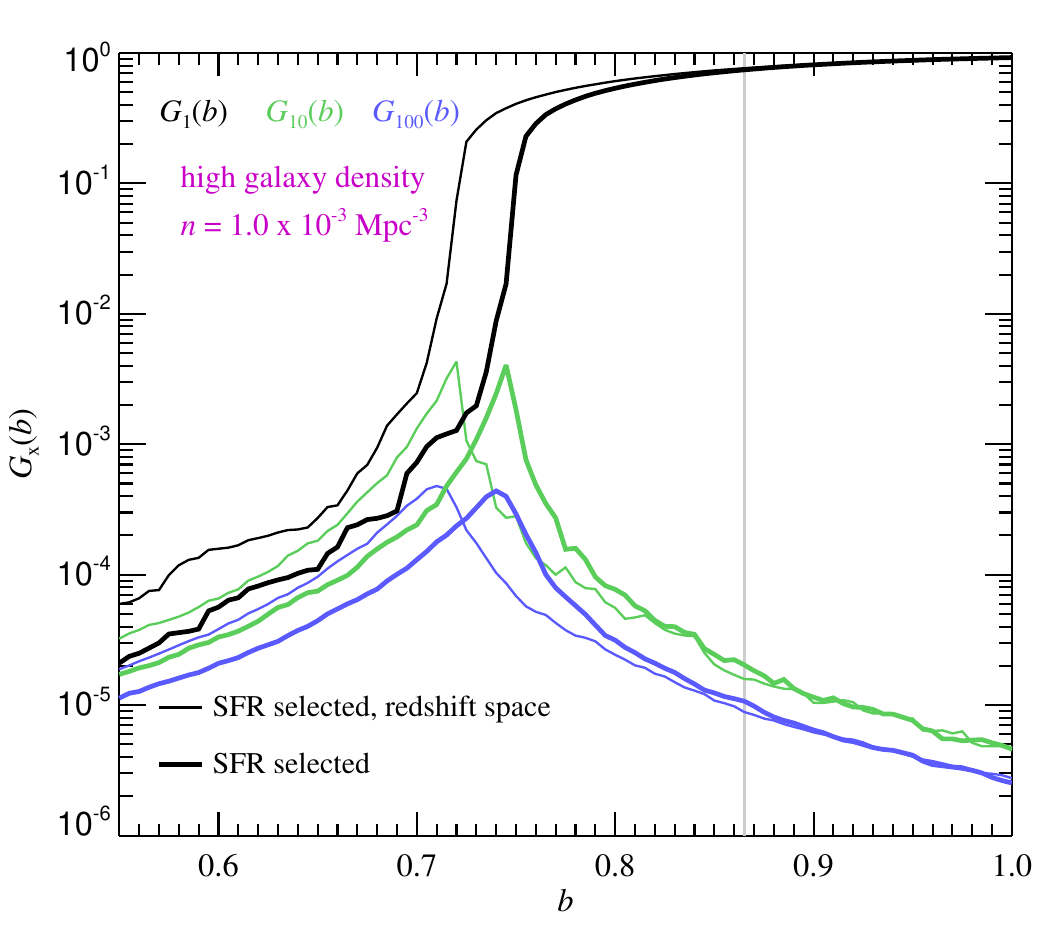}}%
\caption{Differences in percolation statistics between redshift space and real space, for two difference galaxy sample densities, low (left panel) or high (right panel). In this case, the selection criterion was based on star formation rate, but the size of the induced differences from redshift space distortions is similar for selection criteria. For reference, the vertical grey lines show the location of the percolation transition for a random particle distribution.
\label{FigRedshiftSpace}}
\end{figure*}

In Figure~\ref{FigRedshiftSpace} we compare the percolation statistics for the SFR-selected galaxy sample in redshift space to the results reported earlier in real space. We carry out this comparison separately for our lowest (left panel) and our highest (right panel) galaxy density. For conciseness we omit reporting results for other galaxy densities and selection criteria, which produce qualitatively similar results.

Interestingly, the results of Fig.~\ref{FigRedshiftSpace} show that galaxies in redshift space  tend to percolate significantly earlier than in real space, and that this effect is more pronounced for our high galaxy number density. Note that the corresponding shifts are similar to -- or even considerably higher -- than the differences in percolation threshold seen between SFR- and halo-selected samples in real space (Fig.~\ref{FigCompMhalpSFR}). Again, this is a fairly sobering result for the discriminative power of percolation statistics as a practical tool in cosmology. Note also that our results do not confirm earlier findings by \citet{Zhang2018} who reported an insensitivity of percolation statistics to redshift space distortions. However, their galaxy sampling density is comparatively low, a regime where we also find only a small impact of redshift space distortions.

\section{Discussion}
\label{SecDiscussion}

Our results have confirmed that the percolation transition marks an interesting collective feature of the large-scale structure. In particular, we have demonstrated that at low redshift the percolation threshold is markedly different from the one expected for a random point distribution. We have also seen that the percolation statistics is sensitive to the redshift evolution of cosmic structure, and can even detect the difference in the neutrino distribution between neutrino rest masses of $100\,{\rm meV}$ and $300\,{\rm  meV}$. At face value, this seems to suggest that percolation statistics is a potentially very powerful and promising tool for the analysis of cosmic large-scale structure.

We have also collected substantial evidence that unfortunately tarnish this assessment. First of all, we found that numerical resolution has a significant influence on the percolation statistics of the dark matter distribution, for {\it identical} cosmic large-scale structure. This suggests that percolation statistics is perturbed by shot-noise to a significant and non-negligible degree.

When we analysed galaxy mock catalogues instead, this influence became even more apparent. We found that the results of percolation statistics are often driven more strongly by the different galaxy sample densities that we considered than, for example, by the different galaxy selection criteria. The latter, however, can be easily distinguished by conventional two-point clustering statistics for the case of SFR- and halo-selection, whereas this turns out to be hard using percolation statistics. 

A central issue with continuum percolation statistics is that its results are polluted by shot noise in a way that cannot be easily suppressed. Note that this is different from conventional two-point statistics. Here the power spectrum has a constant additive component due to shot noise that can simply be subtracted if desired (and importantly, the amplitude of this is precisely known for a given galaxy number density), and the complementary auto-correlation function is only distorted at the origin, something that can be trivially ignored and causes no harm. Unfortunately, the situation is not nearly as clean for the percolation statistics, where the shot noise distorts the results everywhere in a way that is quantitatively not well understood.

One possibility to try to combat shot noise would be to map the point or galaxy  distribution to a grid and apply a smoothing kernel of a fixed or potentially even adaptive size. Note that the use of a grid necessitates smoothing, otherwise one would be extremely sensitive to the grid resolution. This approach has been tried in the literature, also for other global topological like the genus and sometimes also with adaptive smoothing \citep[e.g.][]{Springel1998}. However the problems due to shot noise persist also in these approaches, so we
have not pursued this option here.

Furthermore, it turns out that redshift space distortions have a quite substantial effect on the percolation statistics, lowering the percolation threshold, probably simply due to the finger-of-god effect. It is thus unfortunately not true that redshift space distortions can be neglected for percolation statistics, at least not at the comparatively high galaxy sample densities that we have considered here.

\begin{figure}
\resizebox{8.5cm}{!}{\includegraphics{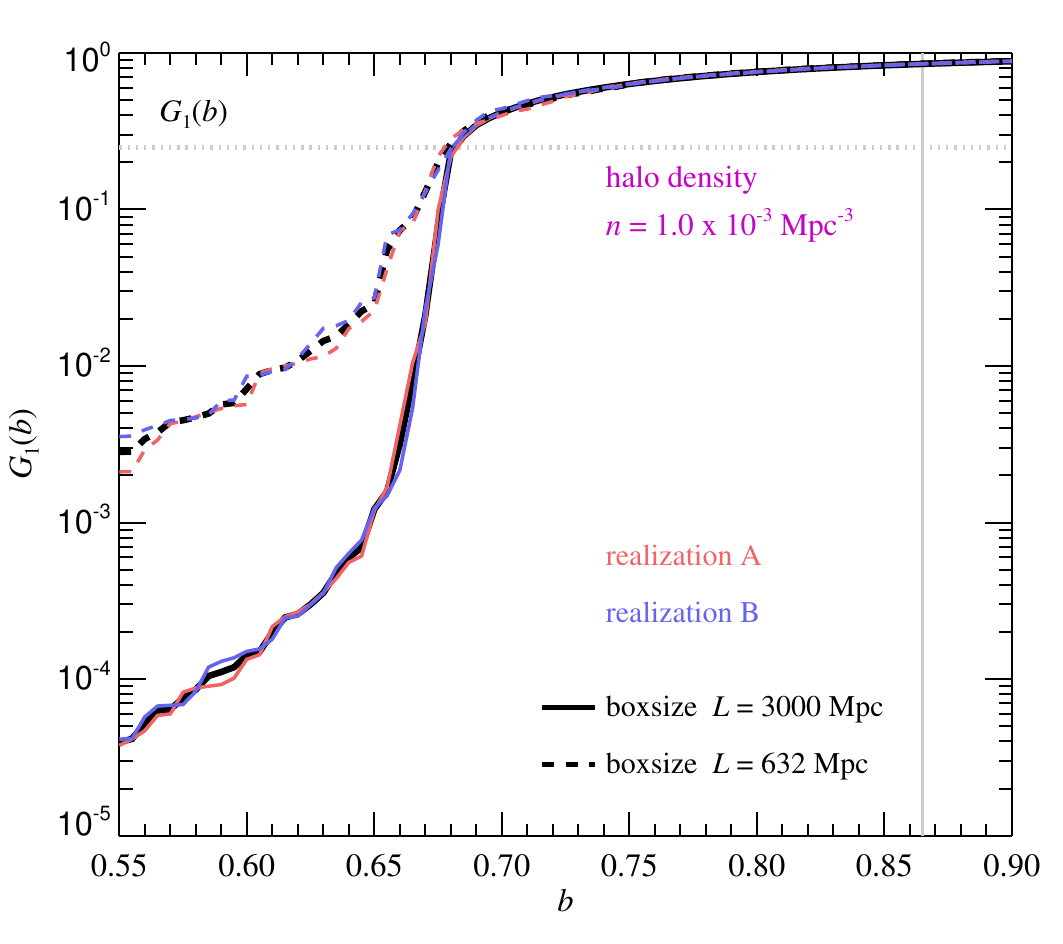}}%
\caption{Percolation statistics for halo samples of constant space density equal to $n=1.0\times 10^{-3}\,{\rm Mpc}^{-3}$, selected by virial mass at $z=0$, in four different simulations with the same cosmological model. We compare two box sizes, equal to $L=3000\,{\rm Mpc}$ and $L=632\,{\rm Mpc}$, each carried out with two paired realizations A and B, as labelled. The thick lines give the average of the two runs for the corresponding box size, while the thin lines show the individual simulations. Clearly, the large box produces results that are hardly affected by cosmic variance at all, and even the small box shows only small differences between the realizations A and B. Note, however, that prior to the percolation statistics, the results for $G_1(b)$ do depend not only on the sampling density but also on the box size. However, if one defines the percolation transition through, for example $G_1(b)=0.25$ (dotted line), nearly invariant results for the percolation threshold are obtained.
\label{FigCosmicVariance}}
\end{figure}

Another interesting question is whether cosmic variance is a significant issue for our results, given that we have thus far analyzed individual simulation realizations of specific cosmological models. Note, however, that our volumes have been comparatively large, especially for our galaxy catalogues, which were constructed for a simulation box 3000 Mpc on side. This is usually considered to be large enough to give results that are fully representative for the underlying cosmological model. A case in point is the good reproduction of the baryonic acoustic oscillation feature in the two-point correlation function seen in Figure~\ref{FigTwoPoint}. But to firm this up more, we consider in Figure~\ref{FigCosmicVariance} the percolation statistics for both of the two simulations we have carried out in the MTNG project for the $3000\,{\rm Mpc}$ boxsize. These realizations differ in the sign of their initial linear displacement field, so that their cosmic web ends up being complementary to each other, in the sense that clusters tend to become voids and vice versa in the evolved simulations. In addition, we also include results for two simulations in a much smaller box of $632\,{\rm Mpc}$ on a side, at the same mass resolution and using the same cosmological model (i.e.~the particle number and volume of these simulations are more than 100 times smaller than for the big boxes). Again, we consider two realizations of this smaller box. For definiteness, we consider the most massive halos selected by virial mass with space density $n=1.0\times 10^{-3}\,{\rm Mpc}^{-3}$ in all the cases.

As Figure~\ref{FigCosmicVariance} shows, the results of the two simulation realizations are very similar in both cases, with the smaller boxes showing slightly larger cosmic variance scatter whereas this is essentially negligible for the big box.  Note, however, that prior to the percolation transition, the percolation curve does depend on the box size (or equivalently on the total number of halos) even though the sample density is the same -- this is the same effect that we already noticed for random point distributions. However, the percolation transition occurs at the same place for both box sizes, and if one defines this implicitly through, for example, $G_1(b)=0.25$ one obtains results for the percolation threshold that are independent of box size to good accuracy.

Our results thus support the cautionary warnings of \citet{Dekel1985} in early work on percolation statistics. A meaningful application of percolation statistics as a cosmological test seems only be possible if the number densities and volumes of observational and theoretical samples are matched. In addition, one needs to consistently work in redshift or real space. Then, however, percolation can still provide a useful test for checking whether the large-scale geometry of the cosmic web is consistent or not. In this function it can complement other topological and morphological measures that are sensitive to higher-order
correlation functions.

\section{Conclusions}
\label{SecConclusions}

Motivated by the large volumes and high resolution of the most recent generation of cosmological simulations, we have revisited percolation statistics as a potentially powerful tool to quantify the geometry of large-scale structure. This statistics goes beyond standard measures of clustering such as the two-point autocorrelation function, and as such probes aspects of the galaxy and matter distribution that are sensitive to global morphological properties of the cosmic web that could be different in alternative cosmologies. Consistency between observed and theoretically predicted percolation statistics can therefore be viewed as a complementary test for the $\Lambda$CDM paradigm.

We have here investigated how the percolation statistics behaves when applied to the dark matter distribution of modern cosmological N-body simulations, as well as to physically based large galaxy mock catalogues in volumes as large as $(3000\, {\rm  Mpc})^3$. Our primary findings can be summarized as follows:
\begin{itemize}
\item The critical percolation threshold for a random particle distribution is invariant with the particle density. The sizes of the largest structures in the regime below percolation are independent of particle density, conversely,  the mass fraction contained in the largest group depends on particle density.

\item In difference to the above, for a cosmological N-body simulations carried out with matching phases at different resolution, the percolation curves are invariant with resolution below the percolation threshold when expressed in terms of relative mass, reflecting the convergence of the masses of the largest groups. However, the percolation threshold itself does tend to drift towards smaller values with higher numerical resolution. We interpret this lack of convergence as a reflection of the global influence of shot noise on the percolation transition, which in this statistics cannot be easily suppressed or subtracted.
  
\item We note that at $z=0$, the percolation threshold and the percolation curves of the dark matter are dramatically different from a random distribution of particles. Interestingly, even for the neutrino component in simulations that model either neutrinos with $100\,{\rm  meV}$ or $300\, {\rm meV}$ rest mass a large difference is found, and in fact, the two neutrino masses can be easily told apart by percolation statistics despite a substantial influence of shot noise on their distributions.

\item We find a strong evolution of the percolation statistics of the dark matter distribution with redshift. The increased clustering towards lower redshift shifts the percolation threshold towards lower values, and increases the values of the percolation functions $G_x(b)$ we defined. At very high redshift, the cold dark matter percolates at critical values that are higher than the one for a Poisson distribution, reflecting the character of  the initial particle distribution as an only slightly perturbed version of the unperturbed Lagrangian distribution. In the subsequent evolution towards lower percolation thresholds, the crossing of the Poissonian value occurs approximately at redshifts around $z\approx 6 $.

\item For large galaxy mock catalogues, we find that the percolation transition depends quite sensitively on the sample density, independent of the detailed galaxy selection criterion. In fact, this dependence is in general stronger than the differences induced by different selection criteria, such as SFR-selection versus stellar mass selection. This suggests that a meaningful discrimination between different geometric galaxy distributions can only be obtained when controlling for an equal number density of the tracers.

\item Galaxy distributions with equal number density but different two-point autocorrelation functions may nevertheless have extremely similar percolation statistics (see in particular the result in Fig.~\ref{FigCompMhalpSFR}, right panel).

\item Redshift space distortions tend to  lower the percolation threshold at all galaxy densities. The differences between real space and redshift space tend to be sizable, and typically larger than variations induced by different galaxy selection criteria.
  
\end{itemize}

Overall, one has to concede that percolation statistics is a cosmological measure that is difficult to interpret due to its dependence on nuisance parameters such as sample density, and its sensitivity to systematic influences such as redshift space distortions. In addition, its discriminative power with respect to different galaxy selection criteria, which in part can be easily told apart by two-point statistics, is arguably disappointing. This may indicate that this measure is of limited practical utility as cosmological probe, and that rather other global measures of large scale structure geometry should be preferred. However, one should not forget that we only tested different tracers of a single $\Lambda$CDM cosmology in this work. There remains the possibility that percolation statistics can well detect differences induced by certain non-standard cosmological models (e.g.~non-Gaussian models, modified gravity scenarios, etc.), so at the very least it remains a consistency check that is worthwhile to be carried out.

\section*{Acknowledgements}

We thank the anonymous referee for constructive and insightful comments that helped to improve the paper. For carrying out the MillenniumTNG simulations, the authors gratefully acknowledge the Gauss Centre for Supercomputing (GCS) for providing computing time on the GCS Supercomputer SuperMUC-NG at the Leibniz Supercomputing Centre (LRZ) in Garching, Germany, under project pn34mo. The  work also used the DiRAC@Durham facility managed by the Institute for Computational Cosmology on behalf of the STFC DiRAC HPC Facility, with equipment funded by BEIS capital funding via STFC capital grants ST/K00042X/1, ST/P002293/1, ST/R002371/1 and ST/S002502/1, Durham University and STFC operations grant ST/R000832/1.
ER is supported by NKFIH-OTKA grant K-142534 from the National Research, Development and Innovation Office, Hungary.
SB is supported by the UK Research and Innovation (UKRI) Future Leaders Fellowship [grant number MR/V023381/1].
CH-A acknowledges support from the Excellence Cluster ORIGINS which is funded by the Deutsche Forschungsgemeinschaft (DFG, German Research Foundation) under Germany’s Excellence Strategy – EXC-2094 – 390783311.

\section*{Data Availability}

The simulations of the MillenniumTNG project are foreseen to be made fully publicly available in 2024 at the following address:
\url{https://www.mtng-project.org}.

%%%%%%%%%%%%%%%%%%%% REFERENCES %%%%%%%%%%%%%%%%%%

% The best way to enter references is to use BibTeX:

%\bibliographystyle{mnras}
\bibliography{main}

\begin{thebibliography}{}
\expandafter\ifx\csname natexlab\endcsname\relax\def\natexlab#1{#1}\fi
\providecommand{\url}[1]{\href{#1}{#1}}
\providecommand{\dodoi}[1]{doi:~\href{http://doi.org/#1}{\nolinkurl{#1}}}
\providecommand{\doeprint}[1]{\href{http://ascl.net/#1}{\nolinkurl{http://ascl.net/#1}}}
\providecommand{\doarXiv}[1]{\href{https://arxiv.org/abs/#1}{\nolinkurl{https://arxiv.org/abs/#1}}}

\bibitem[{{Angulo} \& {Pontzen}(2016)}]{Angulo2016}
{Angulo}, R.~E., \& {Pontzen}, A. 2016, \mnras, 462, L1,
  \dodoi{10.1093/mnrasl/slw098}

\bibitem[{{Angulo} {et~al.}(2014){Angulo}, {White}, {Springel}, \&
  {Henriques}}]{Angulo2014}
{Angulo}, R.~E., {White}, S.~D.~M., {Springel}, V., \& {Henriques}, B. 2014,
  \mnras, 442, 2131, \dodoi{10.1093/mnras/stu905}

\bibitem[{{Banerjee} \& {Abel}(2021)}]{Banerjee2021}
{Banerjee}, A., \& {Abel}, T. 2021, \mnras, 500, 5479,
  \dodoi{10.1093/mnras/staa3604}

\bibitem[{{Barrera} {et~al.}(2023){Barrera}, {Springel}, {White},
  {Hern{\'a}ndez-Aguayo}, {Hernquist}, {Frenk}, {Pakmor}, {Ferlito},
  {Hadzhiyska}, {Delgado}, {Kannan}, \& {Bose}}]{Barrera2023}
{Barrera}, M., {Springel}, V., {White}, S. D.~M., {et~al.} 2023, \mnras, 525,
  6312, \dodoi{10.1093/mnras/stad2688}

\bibitem[{{Berlind} {et~al.}(2006){Berlind}, {Frieman}, {Weinberg}, {Blanton},
  {Warren}, {Abazajian}, {Scranton}, {Hogg}, {Scoccimarro}, {Bahcall},
  {Brinkmann}, {Gott}, {Kleinman}, {Krzesinski}, {Lee}, {Miller}, {Nitta},
  {Schneider}, {Tucker}, {Zehavi}, \& {SDSS Collaboration}}]{Berlind2006}
{Berlind}, A.~A., {Frieman}, J., {Weinberg}, D.~H., {et~al.} 2006, \apjs, 167,
  1, \dodoi{10.1086/508170}

\bibitem[{{Bharadwaj} {et~al.}(2000){Bharadwaj}, {Sahni}, {Sathyaprakash},
  {Shandarin}, \& {Yess}}]{Bharadwaj2000}
{Bharadwaj}, S., {Sahni}, V., {Sathyaprakash}, B.~S., {Shandarin}, S.~F., \&
  {Yess}, C. 2000, \apj, 528, 21, \dodoi{10.1086/308163}

\bibitem[{{Boerner} \& {Mo}(1989)}]{Boerner1989}
{Boerner}, G., \& {Mo}, H. 1989, \aap, 224, 1

\bibitem[{{Bond} {et~al.}(1996){Bond}, {Kofman}, \& {Pogosyan}}]{Bond1996}
{Bond}, J.~R., {Kofman}, L., \& {Pogosyan}, D. 1996, \nat, 380, 603,
  \dodoi{10.1038/380603a0}

\bibitem[{{Bose} {et~al.}(2023){Bose}, {Hadzhiyska}, {Barrera}, {Delgado},
  {Ferlito}, {Frenk}, {Hern{\'a}ndez-Aguayo}, {Hernquist}, {Kannan}, {Pakmor},
  {Springel}, \& {White}}]{Bose2023}
{Bose}, S., {Hadzhiyska}, B., {Barrera}, M., {et~al.} 2023, \mnras, 524, 2579,
  \dodoi{10.1093/mnras/stad1097}

\bibitem[{{Busch} \& {White}(2020)}]{Busch2020}
{Busch}, P., \& {White}, S. D.~M. 2020, \mnras, 493, 5693,
  \dodoi{10.1093/mnras/staa572}

\bibitem[{{Cautun} {et~al.}(2014){Cautun}, {van de Weygaert}, {Jones}, \&
  {Frenk}}]{Cautun2014}
{Cautun}, M., {van de Weygaert}, R., {Jones}, B. J.~T., \& {Frenk}, C.~S. 2014,
  \mnras, 441, 2923, \dodoi{10.1093/mnras/stu768}

\bibitem[{{Contreras} {et~al.}(2023){Contreras}, {Angulo}, {Springel}, {White},
  {Hadzhiyska}, {Hernquist}, {Pakmor}, {Kannan}, {Hern{\'a}ndez-Aguayo},
  {Barrera}, {Ferlito}, {Delgado}, {Bose}, \& {Frenk}}]{Contreras2023}
{Contreras}, S., {Angulo}, R.~E., {Springel}, V., {et~al.} 2023, \mnras, 524,
  2489, \dodoi{10.1093/mnras/stac3699}

\bibitem[{{Davis} {et~al.}(1985){Davis}, {Efstathiou}, {Frenk}, \&
  {White}}]{Davis1985}
{Davis}, M., {Efstathiou}, G., {Frenk}, C.~S., \& {White}, S.~D.~M. 1985, \apj,
  292, 371, \dodoi{10.1086/163168}

\bibitem[{{Dekel} \& {West}(1985)}]{Dekel1985}
{Dekel}, A., \& {West}, M.~J. 1985, \apj, 288, 411, \dodoi{10.1086/162806}

\bibitem[{{Delgado} {et~al.}(2023){Delgado}, {Hadzhiyska}, {Bose}, {Springel},
  {Hernquist}, {Barrera}, {Pakmor}, {Ferlito}, {Kannan},
  {Hern{\'a}ndez-Aguayo}, {White}, \& {Frenk}}]{Delgado2023}
{Delgado}, A.~M., {Hadzhiyska}, B., {Bose}, S., {et~al.} 2023, \mnras, 523,
  5899, \dodoi{10.1093/mnras/stad1781}

\bibitem[{{Einasto} {et~al.}(1984){Einasto}, {Klypin}, {Saar}, \&
  {Shandarin}}]{Einasto1984}
{Einasto}, J., {Klypin}, A.~A., {Saar}, E., \& {Shandarin}, S.~F. 1984, \mnras,
  206, 529, \dodoi{10.1093/mnras/206.3.529}

\bibitem[{{Einasto} {et~al.}(2018){Einasto}, {Suhhonenko}, {Liivam{\"a}gi}, \&
  {Einasto}}]{Einasto2018}
{Einasto}, J., {Suhhonenko}, I., {Liivam{\"a}gi}, L.~J., \& {Einasto}, M. 2018,
  \aap, 616, A141, \dodoi{10.1051/0004-6361/201833011}

\bibitem[{{Elbers} {et~al.}(2021){Elbers}, {Frenk}, {Jenkins}, {Li}, \&
  {Pascoli}}]{Elbers2021}
{Elbers}, W., {Frenk}, C.~S., {Jenkins}, A., {Li}, B., \& {Pascoli}, S. 2021,
  \mnras, 507, 2614, \dodoi{10.1093/mnras/stab2260}

\bibitem[{{Ferlito} {et~al.}(2023){Ferlito}, {Springel}, {Davies},
  {Hern{\'a}ndez-Aguayo}, {Pakmor}, {Barrera}, {White}, {Delgado},
  {Hadzhiyska}, {Hernquist}, {Kannan}, {Bose}, \& {Frenk}}]{Ferlito2023}
{Ferlito}, F., {Springel}, V., {Davies}, C.~T., {et~al.} 2023, \mnras, 524,
  5591, \dodoi{10.1093/mnras/stad2205}

\bibitem[{{Hadzhiyska} {et~al.}(2023{\natexlab{a}}){Hadzhiyska}, {Hernquist},
  {Eisenstein}, {Delgado}, {Bose}, {Kannan}, {Pakmor}, {Springel}, {Contreras},
  {Barrera}, {Ferlito}, {Hern{\'a}ndez-Aguayo}, {White}, \&
  {Frenk}}]{Hadzhiyska2023a}
{Hadzhiyska}, B., {Hernquist}, L., {Eisenstein}, D., {et~al.}
  2023{\natexlab{a}}, \mnras, 524, 2524, \dodoi{10.1093/mnras/stad279}

\bibitem[{{Hadzhiyska} {et~al.}(2023{\natexlab{b}}){Hadzhiyska}, {Eisenstein},
  {Hernquist}, {Pakmor}, {Bose}, {Delgado}, {Contreras}, {Kannan}, {White},
  {Springel}, {Frenk}, {Hern{\'a}ndez-Aguayo}, {Barrera}, \&
  {Monica}}]{Hadzhiyska2023b}
{Hadzhiyska}, B., {Eisenstein}, D., {Hernquist}, L., {et~al.}
  2023{\natexlab{b}}, \mnras, 524, 2507, \dodoi{10.1093/mnras/stad731}

\bibitem[{{Hern{\'a}ndez-Aguayo} {et~al.}(2023){Hern{\'a}ndez-Aguayo},
  {Springel}, {Pakmor}, {Barrera}, {Ferlito}, {White}, {Hernquist},
  {Hadzhiyska}, {Delgado}, {Kannan}, {Bose}, \& {Frenk}}]{Aguayo2023}
{Hern{\'a}ndez-Aguayo}, C., {Springel}, V., {Pakmor}, R., {et~al.} 2023,
  \mnras, 524, 2556, \dodoi{10.1093/mnras/stad1657}

\bibitem[{{Kannan} {et~al.}(2023){Kannan}, {Springel}, {Hernquist}, {Pakmor},
  {Delgado}, {Hadzhiyska}, {Hern{\'a}ndez-Aguayo}, {Barrera}, {Ferlito},
  {Bose}, {White}, {Frenk}, {Smith}, \& {Garaldi}}]{Kannan2023}
{Kannan}, R., {Springel}, V., {Hernquist}, L., {et~al.} 2023, \mnras, 524,
  2594, \dodoi{10.1093/mnras/stac3743}

\bibitem[{{Klypin} \& {Shandarin}(1993)}]{Klypin1993}
{Klypin}, A., \& {Shandarin}, S.~F. 1993, \apj, 413, 48, \dodoi{10.1086/172975}

\bibitem[{{Mecke} {et~al.}(1994){Mecke}, {Buchert}, \& {Wagner}}]{Mecke1994}
{Mecke}, K.~R., {Buchert}, T., \& {Wagner}, H. 1994, \aap, 288, 697,
  \dodoi{10.48550/arXiv.astro-ph/9312028}

\bibitem[{{Mo} \& {White}(1996)}]{Mo1996}
{Mo}, H.~J., \& {White}, S.~D.~M. 1996, \mnras, 282, 347,
  \dodoi{10.1093/mnras/282.2.347}

\bibitem[{{Pakmor} {et~al.}(2023){Pakmor}, {Springel}, {Coles}, {Guillet},
  {Pfrommer}, {Bose}, {Barrera}, {Delgado}, {Ferlito}, {Frenk}, {Hadzhiyska},
  {Hern{\'a}ndez-Aguayo}, {Hernquist}, {Kannan}, \& {White}}]{Pakmor2023}
{Pakmor}, R., {Springel}, V., {Coles}, J.~P., {et~al.} 2023, \mnras, 524, 2539,
  \dodoi{10.1093/mnras/stac3620}

\bibitem[{{Pandey} \& {Bharadwaj}(2005)}]{Pandey2005}
{Pandey}, B., \& {Bharadwaj}, S. 2005, \mnras, 357, 1068,
  \dodoi{10.1111/j.1365-2966.2005.08726.x}

\bibitem[{{Peebles}(1974)}]{Peebles1974}
{Peebles}, P.~J.~E. 1974, \apjl, 189, L51, \dodoi{10.1086/181462}

\bibitem[{{Postman} {et~al.}(1989){Postman}, {Spergel}, {Sutin}, \&
  {Juszkiewicz}}]{Postman1989}
{Postman}, M., {Spergel}, D.~N., {Sutin}, B., \& {Juszkiewicz}, R. 1989, \apj,
  346, 588, \dodoi{10.1086/168040}

\bibitem[{{Pranav} {et~al.}(2017){Pranav}, {Edelsbrunner}, {van de Weygaert},
  {Vegter}, {Kerber}, {Jones}, \& {Wintraecken}}]{Pranav2017}
{Pranav}, P., {Edelsbrunner}, H., {van de Weygaert}, R., {et~al.} 2017, \mnras,
  465, 4281, \dodoi{10.1093/mnras/stw2862}

\bibitem[{{Sathyaprakash} {et~al.}(1998){Sathyaprakash}, {Sahni}, {Shandarin},
  \& {Fisher}}]{Sathyaprakash1998}
{Sathyaprakash}, B.~S., {Sahni}, V., {Shandarin}, S., \& {Fisher}, K.~B. 1998,
  \apjl, 507, L109, \dodoi{10.1086/311689}

\bibitem[{{Shandarin}(1983)}]{Shandarin1983}
{Shandarin}, S.~F. 1983, Soviet Astronomy Letters, 9, 104

\bibitem[{Shklovskii \& Efros(1984)}]{Shklovskii1984}
Shklovskii, B.~I., \& Efros, A.~L. 1984, Electronic Properties of Doped
  Semiconductors (Springer-Verlag Berlin Heidelberg),
  \dodoi{10.1007/978-3-662-02403-4}

\bibitem[{{Sousbie}(2011)}]{Sousbie2011}
{Sousbie}, T. 2011, \mnras, 414, 350, \dodoi{10.1111/j.1365-2966.2011.18394.x}

\bibitem[{{Springel} {et~al.}(2021){Springel}, {Pakmor}, {Zier}, \&
  {Reinecke}}]{Springel2021}
{Springel}, V., {Pakmor}, R., {Zier}, O., \& {Reinecke}, M. 2021, \mnras, 506,
  2871, \dodoi{10.1093/mnras/stab1855}

\bibitem[{{Springel} {et~al.}(1998){Springel}, {White}, {Colberg}, {Couchman},
  {Efstathiou}, {Frenk}, {Jenkins}, {Pearce}, {Nelson}, {Peacock}, \&
  {Thomas}}]{Springel1998}
{Springel}, V., {White}, S. D.~M., {Colberg}, J.~M., {et~al.} 1998, \mnras,
  298, 1169, \dodoi{10.1046/j.1365-8711.1998.01710.x}

\bibitem[{{Springel} {et~al.}(2005){Springel}, {White}, {Jenkins}, {Frenk},
  {Yoshida}, {Gao}, {Navarro}, {Thacker}, {Croton}, {Helly}, {Peacock}, {Cole},
  {Thomas}, {Couchman}, {Evrard}, {Colberg}, \& {Pearce}}]{Springel2005}
{Springel}, V., {White}, S. D.~M., {Jenkins}, A., {et~al.} 2005, \nat, 435,
  629, \dodoi{10.1038/nature03597}

\bibitem[{{Stauffer}(1979)}]{Stauffer1979}
{Stauffer}, D. 1979, \physrep, 54, 1, \dodoi{10.1016/0370-1573(79)90060-7}

\bibitem[{{White}(1979)}]{White1979}
{White}, S.~D.~M. 1979, \mnras, 186, 145, \dodoi{10.1093/mnras/186.2.145}

\bibitem[{{Yess} \& {Shandarin}(1996)}]{Yess1996}
{Yess}, C., \& {Shandarin}, S.~F. 1996, \apj, 465, 2, \dodoi{10.1086/177397}

\bibitem[{{Yess} {et~al.}(1997){Yess}, {Shandarin}, \& {Fisher}}]{Yess1997}
{Yess}, C., {Shandarin}, S.~F., \& {Fisher}, K.~B. 1997, \apj, 474, 553,
  \dodoi{10.1086/303496}

\bibitem[{{Zeldovich} {et~al.}(1982){Zeldovich}, {Einasto}, \&
  {Shandarin}}]{Zeldovich1982}
{Zeldovich}, I.~B., {Einasto}, J., \& {Shandarin}, S.~F. 1982, \nat, 300, 407,
  \dodoi{10.1038/300407a0}

\bibitem[{{Zhang} {et~al.}(2018){Zhang}, {Cheng}, \& {Chu}}]{Zhang2018}
{Zhang}, J., {Cheng}, D., \& {Chu}, M.-C. 2018, \prd, 97, 023534,
  \dodoi{10.1103/PhysRevD.97.023534}

\end{thebibliography}

\end{document}